\documentclass[preprint2]{aastex} 
\usepackage{graphicx}
\usepackage{url}
\usepackage{color}

\shorttitle{Eclipse Megamovie}
\shortauthors{Hudson et al.}

\begin{document}

\title{The U.S. Eclipse Megamovie in 2017: \\
\textit{a white paper on a unique outreach event}}

\author{H.~S. Hudson$^{1,2}$, S.~W. McIntosh$^3$, S.~R. Habbal$^4$, J.~M. Pasachoff$^5$, and L. Peticolas$^1$}
\affil{$^1$SSL, UC Berkeley, CA, USA 94720; 
$^2$University of Glasgow; 
$^3$High Altitude Observatory, National Center for Atmospheric Research, P.O. Box 3000, Boulder CO 80307;
$^4$Institute for Astronomy, University of Hawaii, Honolulu HI, USA;
$^5$Williams College, Williamstown MA, USA
}
\email{hhudson@ssl.berkeley.edu}


\begin{abstract}
Totality during the solar eclipse of 2017 traverses the entire breadth of the continental United States, from Oregon to South Carolina.
It thus provides the opportunity to assemble a very large number of images, obtained by amateur observers all along the path, into a continuous record of coronal evolution in time; totality lasts for an hour and a half over the continental U.S.
While we describe this event here as an opportunity for public education and outreach, such a movie -- with very high time resolution and extending to the chromosphere -- will also contain unprecedented information about the physics of the solar corona.
\end{abstract}


\section{The Opportunity}\label{sec:intro}

\begin{figure}[htpb]
\centering
   \includegraphics[width=0.45\textwidth]{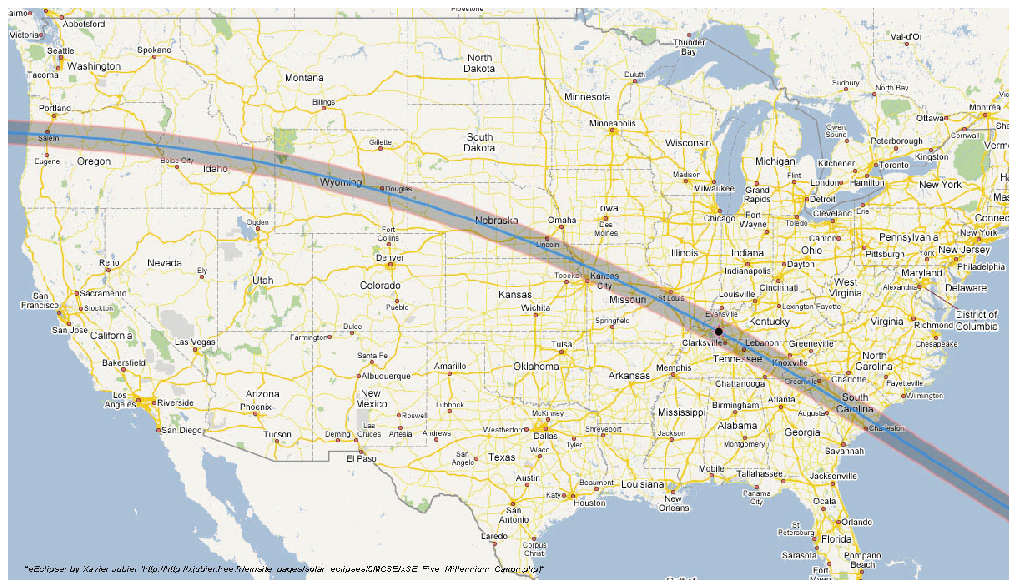}
   \caption{Eclipse map, showing the path of totality, from http://www.eclipse-maps.com
   (see Section~\ref{sec:concl}).
  }  
\label{fig:usmap}
\end{figure}

The 2017 total eclipse of the Sun (\url{http://www.eclipses.info}) begins in the Pacific and ends in the Atlantic, and in between the path of totality crosses as many as 14 of the States
\citep[e.g.,][]{2011arXiv1108.2323H}.
Figure~\ref{fig:usmap} shows the geography.
The point of greatest eclipse will be in Kentucky, near mid-day (18:26:40~UT 21~August 2017), and at that point the duration of totality will be almost three minutes.
Prior to this opportunity, there will be three other total eclipses (all with inferior conditions), the first occurring in November~2012.

Unquestionably this eclipse will attract a great deal of public attention, and we can expect a surge of interest in eclipse photography. 
This white paper describes some current thinking about a ``Megamovie'' program, with a description aimed at the professional community. 
The basic ideas follow the poster presentations by Hudson et al.\footnote{\url{http://sprg.ssl.berkeley.edu/~hhudson/presentations/spd.110613/}} and Habbal et al. at the American Astronomical Society/Solar Physics Division meeting in Las Cruces (June 2011).

\section{The Megamovie}

The basic idea of the Megamovie is to incorporate as many images as possible, provided by a diverse range of observers using standard photographic techniques, into an overview movie. This would show the dynamics of the corona and associated prominence systems at high resolution for an extended period of time. If 10,000 observers each obtained 100 frames, then we would have a million-frame movie; at standard frame rate this would take 12 hours to show, and would thus be a slow-motion representation of coronal evolution. Each participating observer would be able to point with pride to the exact moment of his or her contribution. Of course, with an uncontrollably heterogeneous database, with images acquired by any number of camera types and formats, a substantial effort would be required even to produce a crude product.

On the other hand, many sequences of higher-quality data will surely also be a part of the activity. The organization of the program described here can help to ensure that these will appear via the publication of recommendations and instructions, and via other forms of assistance. These better data will indeed be unique: no eclipse to date has been tracked seamlessly with high time resolution for intervals longer than a single station's duration of totality. Some of these data will succeed in obtaining have spatial resolution. To make a comparison, we note that space-borne coronagraphs have large pixels (11.2$''$ for LASCO/C2, for example) and extremely low cadences.
Good eclipse imagery should have much higher resolution, depending on the instrumentation.

With new capabilities we expect to see many manifestations of dynamic coronal activity, extending far into the corona.

\section{How to deal with the data?}

Composite eclipse images of the very highest quality have now frequently been produced by Miloslav Druckm{\" u}ller and his collaborators (\url{http://www.zam.fme.vutbr.cz/~druck/}). Obtaining such precision of registration and photometric adjustment could not happen automatically, and yet with such a flood of miscellaneous data arriving within a short time interval, the Megamovie project will have to develop ways to handle input as a pipeline data product. Indeed, a Megamovie data portal would need to be able to accept of order 10 TB of data within a few days following the event, and then pipeline-style software would ideally edit these data into a rough-cut movie suitable for distribution on a short time scale.
Software and facilities to enable this reaction will need to be identified and developed.

\begin{figure}[htpb]
\centering
   \includegraphics[width=0.45\textwidth]{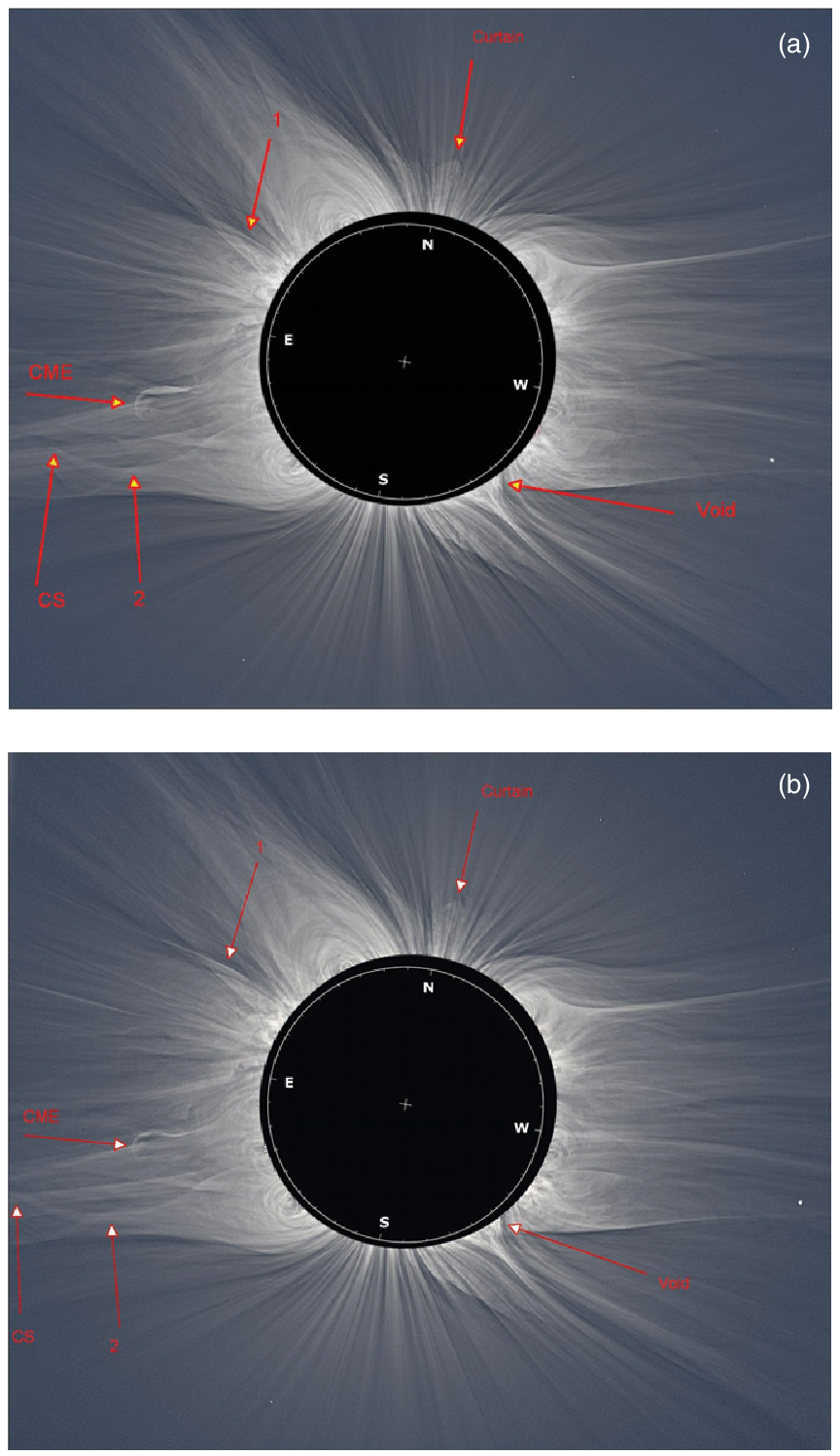}
   \caption{Two views of the 2010 eclipse, separated by almost 1.5 hours \citep[from][]{2011ApJ...734..114P}.
   Note the motion of the feature labeled ``CME'' for example \citep[see also][]{2011ApJ...734..120H}.
  }  
\label{fig:pasachoff}
\end{figure}

The best sequences of images will warrant careful coalignment and image processing. It is likely that some sequences will have extremely high quality and will have captured specific dynamical events in the corona. At the extreme level it may be possible to create a very few composite frames, extended over a long time interval, using techniques such as those of Druckm{\" u}ller. The 2010 total eclipse (Pacific islands and Chile) provided an example of such a comparison, as shown in Figure~\ref{fig:pasachoff} \citep{2011ApJ...734..114P}.

At a minimum, the Megamovie data system must be able to do quasi-real-time image manipulation with a minimum level of astrometric calibration (scale, orientation) and histogram matching, taking into account whatever metadata may be available but not relying upon their correctness (or even their existence). \url{astrometry.net} has developed software that successfully deals with such issues for star fields \citep[e.g.,][and references therein]{2011arXiv1103.6038L}. \cite{2006..............D} and \cite{2009ApJ...706.1605D} discuss some of the techniques for such calibrations.

\section{Outreach Development}
The educational aspect of the Megamovie project, dealing with schools and community groups of all levels, probably needs separate management but close coordination with the program of data acquisition and re-distribution. Many specific topics need to be covered: school programs, partnerships with existing organizations, outreach to minority groups such as Native Americans, programs in national parks, and development of education software, for example.

\section{Solar Research}
Eclipse observations for research purposes now have substantial priority, thanks partly to great improvements in instrument capability (Pasachoff, 2009; Habbal et al., 2011).
\nocite{2009Natur.459..789P}
Recent interesting experiments have included direct temperature measurements via the spectrum of the K-corona, and spectroscopic analysis of many forbidden lines of highly ionized Fe \citep[e.g.,][]{2011ApJ...734..120H}. 
The particular niche made possible by the Megamovie has not yet been exploited, though: what can we see with a continuous high-resolution movie? 
This long time series of observations will also monitor variations in the low corona, a domain not accessible to the space-based coronagraphs.
The database assembled from amateur imagery therefore will contain items of great scientific value, owing simply to the uniqueness of the approach.

Figure~\ref{fig:pasachoff} shows evolving features in a ``movie'' of very high image quality but with only two frames widely separated in time.
This 2010 eclipse has also produced data now undergoing astrometric reductions for gravitational lensing (see next section).

\section{Relativity}
\begin{figure}[htpb]
\centering
   \includegraphics[width=0.45\textwidth]{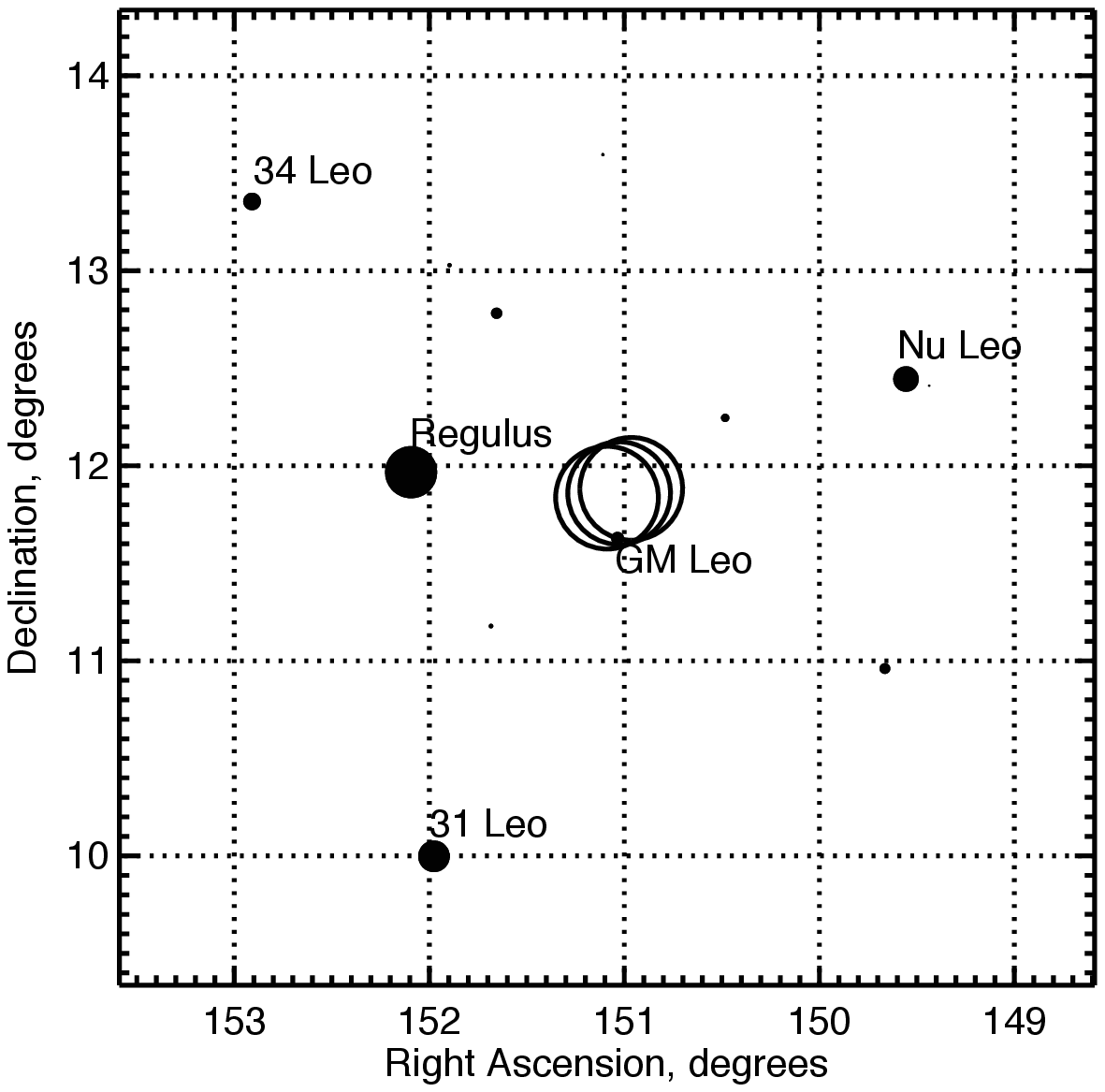}
   \caption{Background star field for the 2017 eclipse. The three circles show the solar disk across the duration of totality in the continental U.S.
  }  
\label{fig:starmap_2017}
\end{figure}

So far as we are aware, nobody has successfully repeated the Eddington experiment \citep{1920RSPTA.220..291D} on the gravitational deflection of starlight, via the use of modern CCD cameras \citep[e.g.,][]{2006spse.conf...55D}.
It is a difficult experiment, but the deflections are not small (1.75$''/r$, where~$r$ is the elongation in solar radii). 
Nowadays radio techniques are far superior \citep[e.g.,][]{2010IAUS..261..291F}, and so this classical measurement has little scientific value; nevertheless the possibility of checking Einstein using bright stars and modern mega-pixel cameras will probably appeal to many participants in the eclipse program. 
We have noted the existence of  \url{astrometry.net}.
This automated software can calibrate almost any stellar image astrometrically.
We  suggest testing this software approach on images that already exist or that can be obtained in earlier eclipses, such as that of 2012 in northern Australia. 
The 2017 eclipse field contains the bright star Regulus ($\alpha$ Leo; $m_V = 1.36$), as illustrated in Figure~\ref{fig:starmap_2017}. 
Regulus will deflect 0.74$''$ at the time of greatest eclipse (18:26:40~UT).\footnote{\url{http://sprg.ssl.berkeley.edu/~stephchow/eclipse/webpage/}}

\section{Program development}

Some topics to consider:

\begin{itemize}
\item The Megamovie base organization(s)
\item Advertising
\item Sponsorship
\item Program documentation
\item Site preparation
\item Outreach preparation
\item Outreach operations
\item Communication
\item Training manuals
\item Specifications for standard observing equipment
\item Web interface
\item Web servers
\item Analysis software
\item Smartphone app development
\item ``Eclipse@home'' survey program
\item Data processing
\item Distribution of products
\item Science
\end{itemize}

\section{A Smaller-Scale Field Test?}
The aforementioned November 2012 eclipse, in northern Australia, provides an ideal opportunity to test data the astrometry processing techniques on a diverse set of image types as well as any prototype data acquisition website and pipeline. 
A relatively large scientific meeting is scheduled to coincide with the event (\url{http://moca.monash.edu/eclipse/}) at which an announcement will be made to request the eclipse photographs taken by the attendees. 
Events such as this eclipse will be sought out by the Megamovie project team as a means of assessing the end-to-end efficiency of such a process, and to help reveal problem areas prior to the main event in 2017.

\section{Conclusions}\label{sec:concl}
The Megamovie project presents many opportunities, and we feel that its simple objectives will bring broad attention to educationally important phenomena.
Thus our organization should emphasize links to other related efforts, and indeed there is extensive Web material already available for this specific event, such as 

{\it 
\noindent http://www.eclipses.info;

\noindent http://eclipse.gsfc.nasa.gov/eclipse.html;

\noindent http://xjubier.free.fr/en/site\_pages/

\noindent  ~~SolarEclipsesGoogleMaps.html; 
 
\noindent http://www.eclipse-maps.com/;

\noindent http://www.hermit.org/eclipse/2017-08-21/;

\noindent http://www.eclipse2017.org/;

\noindent http://www.astrometry.net;

\noindent http://sprg.ssl.berkeley.edu/~stephchow/

\noindent  ~~eclipse/webpage/
}
\bibliographystyle{apj}
\bibliography{ecl}
\date{\today}

\end{document}